\documentclass[twocolumn,showpacs,prb]{revtex4-1}
\usepackage{color}
\usepackage{graphicx,epsfig,amssymb}
\usepackage{bm}
\begin{document}
\title{\bf Bound fermion states in pinned vortices in the surface states of a superconducting topological insulator: The Majorana bound state} 
\vspace{1.5em}
\author{Haoyun Deng, N. Bonesteel and P. Schlottmann}
\affiliation{Department of Physics, Florida State University, Tallahassee, FL 32306, USA}
\affiliation{National High Magnetic Field Laboratory, Florida State University, Tallahassee, Florida 32310, USA} 
\date{\today}
\begin{abstract}
By analytically solving the Bogoliubov-de Gennes equations we study the fermion bound states at the center of the core of a vortex in a two-dimensional superconductor. We consider three kinds of 2D superconducting models: (a) a standard type II superconductor in the mixed state with low density of vortex lines, (b) a superconductor with strong spin-orbit coupling locking the spin parallel to the momentum and (c) a superconductor with strong spin-orbit coupling locking the spin perpendicular to the momentum. The 2D superconducting states are induced via proximity effect between an $s$-wave superconductor and the surface states of a strong topological insulator. In case (a) the energy gap for the excitations is of order $\Delta_{\infty}^2/(2E_F)$, while for cases (b) and (c) a zero-energy Majorana state arises together with an equally spaced ($\Delta^2_{\infty}/E_F$) sequence of fermion excitations. The spin-momentum locking is key to the formation of the Majorana state. We present analytical expressions for the energy spectrum and the wave functions. 
\end{abstract}
\pacs{71.10.Pm, 03.67.Lx, 74.45.+c, 74.90.+n}
\maketitle

\section{Introduction}

Majorana fermions are unconventional quantum states with non-Abelian statistics and potential for quantum computing.\cite{Beenakker} The idea of storing quantum information in Majorana states originates from Kitaev.\cite{Kitaev} The generation of Majorana bound states at surfaces of strong topological insulators (TI) due to the proximity of an $s$-wave superconductor (S) has been explored by Fu and Kane.\cite{FuKane1,FuKane2} Majorana edge states occur at a junction between a superconductor and a ferromagnet deposited on the surface of a topological insulator.\cite{FuKane2,Beenakker} A Majorana state also arises as a zero-energy bound state at the core of a vortex as a consequence of the strong spin-orbit coupling in the topological insulator.\cite{FuKane1,Beenakker} For a review see Ref. [\onlinecite{Beenakker1}].

In this paper we study the electronic structure of vortices in superconducting surface states with strong spin-orbit coupling. For our analytic calculation we follow the method employed by Caroli, de Gennes and Matricon (CdeGM)\cite{CaroliDGM} for a vortex line in a three-dimensional superconductor. The Bogoliubov-de Gennes equations are solved (i) for small distances $\rho$ (compared to the correlation length $\xi$) from the core of the vortex, where the superconductor order parameter can be neglected, and (ii) for larger distances, still smaller than $\xi$, but where the order parameter needs to be taken into account. These two solutions are matched at an intermediate radius $\rho_c$. If the matching condition is such that it is independent of the value of $\rho_c$, then we have a solution for the entire region of the vortex. This condition as well determines the value of the energy of the bound state inside the vortex core. This way we obtain the entire low-energy spectrum of the bound states, as well as the analytic expression of the corresponding wave functions. In Sect. II we consider the bound fermion states in the vortex core of a 2D superconductor. This model contains no spin-orbit coupling and corresponds to the reduction of the 3D CdeGM calculation to 2D. Consequently the bound states are gapped from the ground state by a small gap of order $\Delta_{\infty}^2/(2E_F)$. Although Sect. II does not contain new information, we believe it is a pedagogically useful basis for the remaining sections.

There are previous studies of bound states in type II superconductor vortices besides Refs. [\onlinecite{CaroliDGM}] and [\onlinecite{CaroliM}]. Based on a generalized Ginzburg-Landau theory, Neumann and Tewordt\cite{Neumann} considered a free-energy functional including terms to the fourth order in $\Delta$ to obtain the electronic structure of a vortex line.  Using the WKBJ approximation the structure of vortex lines in pure superconductors was investigated by Bardeen {\it et al.}\cite{Bardeen} Within the framework of the Bogoliubov-de Gennes theory Gygi and Schl\"uter\cite{Gygi} calculated the spectrum of a type II superconductor vortex and several related properties and successfully compared their results with scanning-tunneling-microscopy experiments on NbSe$_2$. The bound states in a vortex along the BCS-BEC crossover were studied by Simonucci {\it et al.} using the Bogoliubov-de Gennes equations.\cite{Simonucci1,Simonucci2} Finally, Rainer {\it et al.},\cite{Rainer} in the context of high-$T_c$ studied the spectrum of an isolated ``stack'' of pancake vortices in clean layered superconductors and concluded that both, the circular current around the vortex center as well as transport currents through the vortex core are carried by localized states bound to the core.

Studies of vortices with more than one flux quantum (vorticity larger than one) in type II superconductors have also been studied in numerous papers.\cite{Volovik,Salomaa,Berthod,Lasher} The results strongly depend on whether the vorticity is odd or even. Usually a vortex with more than one flux quantum is unstable and splits into vortices with one flux quantum. Therefore in this paper we limit our calculation to a vortex of vorticity one. 

Following Fu and Kane\cite{FuKane1,FuKane2} numerous authors investigated the surface states of a 3D topological insulator with proximity induced $s$-wave superconductivity in different geometries.\cite{DasSarma1,DasSarma2} Depending on the transparency of the TI/S junction there is a reduction of the induced superconducting gap as compared to the parent S gap. The induced gap is energy dependent, but is only a weak function of energy for low-lying excitations, so that we can consider $\Delta$ a constant.  In the dirty limit the proximity effect has been studied using the Eilenberger/Usadel formalism\cite{Hugdal,Loss,Bobkova} and the Majorana state with the Bogoliubov-de Gennes method for a supersymmetric $\sigma$ model.\cite{Feigelman} A generic platform for topological quantum computation using semiconductor heterostructures was presented in Refs. [\onlinecite{DasSarma3}] and [\onlinecite{Mao}]. 

The electronic structure of a vortex in a topological superconductor has been investigated by numerous authors\cite{DasSarma1,Suzuki,Nori,Nori1,Nori2,Nori3,Jackiw,Feigelman} by solving the  Bogoliubov-de Gennes equations. The strong spin-orbit coupling leads to spin-momentum locking and a zero-energy Majorana bound state, as a consequence of the Berry phase. The low-energy excitations are equally spaced by the amount of $\Delta_{\infty}^2/E_F$. Vortex bound states in a proximity-induced topological superconductor on a spherical surface have been studied in Refs. [\onlinecite{Kraus}] and [\onlinecite{Hu}]. In this geometry the boundary of the topological states is closed corresponding to a vortex anti-vortex pair. Following the proposal by Fu and Kane\cite{FuKane1} in Ref. [\onlinecite{DasSarma1}] the excitation gap in a line junction and a trijunction pair linked by a line junction was considered. Durst\cite{Durst} obtained the scattering cross section for quasi-particles with excitation energy greater that $\Delta_{\infty}$ off the vortex states. A possible application of non-Abelian topological order in $s$-wave superfluids of ultracold fermionic atoms has been proposed by Sato {\it et al.}\cite{Sato}

Most of the above calculations for the vortex bound states are numerical and are restricted to energy eigenvalues. The purpose of this paper is to present analytical expressions for the low-energy bound state eigenvalues and eigenfunctions close to the core of an isolated vortex using the CdeGM method. We consider the metallic surface states of a 3D topological insulator with proximity induced $s$-wave superconductivity. We simplify the model by directly introducing the superconducting order parameter into the 2D electron gas, since the more tedious problem consisting of the TI interacting with S has already been solved.\cite{DasSarma1,Nori}  Two models are considered: (a) a superconductor with strong spin-orbit coupling locking the spin parallel to the momentum (Dirac Hamiltonian) in Sect. III and (b) a S with strong spin-orbit interaction coupling the spin perpendicular to the momentum (Rashba coupling) in Sect. IV. In both cases a zero-energy Majorana state is generated, as a consequence of the strong spin-orbit coupling.  Although the two models can be transformed into each other via a unitary transformation (see Appendix), we believe it is pedagogically useful to solve them independently. Conclusions follow in Sect. V. 

\section{Bound fermion states in a vortex of a 2D superconductor}
 
We consider a 2D type II superconductor in the mixed state with the magnetic field slightly above $H_{c1}$, but $H \ll H_{c2}$, so that we can assume there is an isolated vortex at the origin. The superconducting pair potential is given by $\Delta({\bf r}) = \Delta(\rho) e^{-i\theta}$, where $(\rho,\theta)$ are polar coordinates. Here $\Delta(\rho)$ is real, vanishes for $\rho=0$, increases linearly with $\rho$ and saturates at the value $\Delta_{\infty}$ for $\rho$ larger than the coherence length $\xi$.\cite{Abrikosov}

The Bogoliubov equations are linear coupled differential equations determining two functions, $u({\bf r})$ and $v({\bf r})$, constituting a spinor ${\hat \varphi}^T = \bigl(u({\bf r}),v({\bf r})\bigr)$.  For a three-dimensional superconductor these equations have been studied by Caroli {\it et al.}\cite{CaroliDGM,CaroliM} The present calculation is simplified with respect to the 3D one in that the third dimension is suppressed. The phase of the order parameter is eliminated by the gauge transformation $u=e^{-i\theta/2}u'$ and $v=e^{i\theta/2}v'$, or in spinor notation ${\hat \varphi} = \exp(-i {\hat \sigma}_z \theta/2) {\hat \varphi'}$, where ${\hat \sigma_i}$ are Pauli matrices acting on the spinor. Using the same arguments as Caroli {\it et al.}\cite{CaroliDGM} the vector potential and the magnetic field can be neglected for $\rho < \xi$. The solution of the Bogoliubov equations is then of the form ${\hat \varphi'} = \exp(i\mu\theta) {\hat f}(\rho)$, where $\mu$ is half of an odd integer since ${\hat \varphi}$ should be invariant under rotations of multiples of $2\pi$ to yield a single-valued wave function.\cite{CaroliM} The differential equation satisfied by ${\hat f}$ is of second order and given by
\begin{eqnarray}
&&{\hat \sigma}_z \frac{1}{2m} \left[-\frac{d^2 {\hat f}}{d\rho^2} - \frac{1}{\rho}\frac{d {\hat f}}{d \rho} + \left(\mu -\frac{1}{2}{\hat \sigma}_z\right)^2 \frac{{\hat f}}{\rho^2} - k_F^2 {\hat f} \right] \nonumber \\
&& \ \ \ \ \ \ + \Delta {\hat \sigma}_x {\hat f} = E {\hat f} \ \ , \label{Bog}
\end{eqnarray}
where $|\mu| \ll k_F \xi$, $k_F$ is the Fermi momentum and $\hbar$ is set equal to 1.

Since $\Delta(\rho)$ increases linearly with $\rho$ from $\Delta(0)=0$, we may neglect $\Delta(\rho)$ for sufficiently small $\rho$. Eq. (\ref{Bog}) is then diagonal in the spinor components, ${\hat f}^T = (f_+,f_-)$, and the solution can be expressed in terms of Bessel functions
\begin{equation}
f_{\pm}(\rho) = A_{\pm} J_{\mu\mp1/2}[(k_F \pm q)\rho] \ , \label{Bessel} 
\end{equation} 
where $q = E/v_F$ and $A_{\pm}$ are constants. Here we assumed that $q \ll k_F$ and $(k_F^2 + 2mE\sigma_z)^{1/2} = k_F (1\pm2E/v_Fk_F)^{1/2} \sim k_F \pm q$.

On the other hand, for larger $\rho$, but $\rho < \xi$, $\Delta(\rho)$ is still linear in $\rho$ but cannot be neglected. Due to the order parameter the two components of the spinor are now mixed. Following CdeGM\cite{CaroliDGM} the Ansatz for a solution is ${\hat f}(\rho) = H^{(1)}_{\mu}(\rho) {\hat g}(\rho)+c.c.$, where $H^{(1)}_{\mu}$ is the Hankel function of the first kind of order $\mu$ and ${\hat g}(\rho)$ consists of a pair of slowly varying functions (compared to $H^{(1)}_{\mu}(\rho)$), i.e. envelop functions. Inserting the Ansatz for a solution into the differential equation (\ref{Bog}) and using the differential equation satisfied by the Hankel function, we obtain
\begin{widetext}
\begin{equation}
{\hat \sigma}_z \frac{1}{2m} \left[-H^{(1)}_{\mu}\frac{d^2 {\hat g}}{d\rho^2} - \Bigl(2 \frac{d H^{(1)}_{\mu}}{d\rho}+ \frac{H^{(1)}_{\mu}}{\rho}\Bigr) \frac{d{\hat g}}{d\rho} -\mu{\hat \sigma}_z H^{(1)}_{\mu} \frac{\hat g}{\rho^2} + \frac{H^{(1)}_{\mu}}{4} \frac{\hat g}{\rho^2} \right] + \Delta {\hat \sigma}_x H^{(1)}_{\mu} {\hat g} = E H^{(1)}_{\mu} {\hat g} \ \ . \label{Bog1}
\end{equation}
\end{widetext}
Here we use Bardeen {\it et al.}'s\cite{Bardeen} choice for the order of the Hankel function, while CdeGM\cite{CaroliDGM} considered $H^{(1)}_m$ with $m=\sqrt{\mu^2+{\textstyle \frac{1}{4}}}$. Both are viable ways to proceed. Dividing the equation by $H^{(1)}_{\mu}$, neglecting the term with $\frac{d^2 {\hat g}}{d\rho^2}$ (since ${\hat g}$ is a slowly varying envelop) and using the asymptotic expansion of $H^{(1)}_{\mu}(k_F \rho)$ for large argument
\begin{equation}
\frac{1}{H^{(1)}_{\mu}(k_F \rho)} \frac{d H^{(1)}_{\mu}(k_F \rho)}{d\rho} \sim - \frac{1}{2 \rho} + i k_F \ \ , \label{asympt}
\end{equation}
the differential equation for ${\hat g}$ reduces to
\begin{equation}
-i {\hat \sigma}_z v_F \frac{d {\hat g}}{d \rho} + {\hat \sigma}_x \Delta {\hat g} = \Bigl(E + \frac{\mu}{2m\rho^2} - \frac{{\hat \sigma}_z}{8 m \rho^2} \Bigr) {\hat g} \ \ . \label{g1}
\end{equation}
Note that the next order in the expansion in Eq. (\ref{asympt}) yields only negligible contributions. 

The terms on the rhs of Eq. (\ref{g1}) are small compared to those on the lhs and can be considered a perturbation. We write then ${\hat g} = {\hat g}_0 + {\hat g}_1$, where ${\hat g}_0$ satisfies 
\begin{equation}
-i {\hat \sigma}_z v_F \frac{d {\hat g}_0}{d \rho} + {\hat \sigma}_x \Delta {\hat g}_0 = 0 \ \ . \label{g10}
\end{equation}
The solution of this differential equation is 
\begin{eqnarray}
{\hat g}_0(\rho) &=& C \left(\begin{array}{c} 1 \\ -i \end{array} \right) \exp[-K(\rho)] \ \ , \nonumber \\
K(\rho) &=& \frac{1}{v_F} \int_0^{\rho} d\rho' \Delta(\rho') \ \ . \label{g0+K}
\end{eqnarray}
Note that since ${\hat \sigma}_x$ has two eigenvalues, there is a second solution increasing with $\rho$ as $e^{+K(\rho)}$. This solution, however, can be disregarded, since we expect ${\hat g}_0$ to decrease as $\rho$ increases (the bound states are localized in the vortex core).

The first order perturbation correction due to the rhs of Eq. (\ref{g1}) is obtained through
\begin{equation}
-i {\hat \sigma}_z v_F \frac{d {\hat g}_1}{d \rho} + {\hat \sigma}_x \Delta {\hat g}_1 = \Bigl(E + \frac{\mu}{2m\rho^2} - \frac{{\hat \sigma}_z}{8 m \rho^2}\Bigr) {\hat g}_0 \ \ , \label{g11}
\end{equation}
where Eq. (\ref{g0+K}) is inserted on the rhs for ${\hat g}_0$. For ${\hat g}_1$ we choose the Ansatz $g_1^+ = a_+ e^{-K(\rho)}$ and $g_1^- = -i a_- e^{-K(\rho)}$ and obtain coupled differential equations for $a_+$ and $a_-$
\begin{eqnarray}
&&v_F \left[\frac{d a_+}{d\rho}-\frac{dK}{d\rho} a_+\right] + \Delta a_- = i C \Bigl(E + \frac{\mu-{\textstyle \frac{1}{4}}}{2m\rho^2}\Bigr) \ ,  \nonumber \\
&&v_F \left[\frac{d a_-}{d\rho}-\frac{dK}{d\rho} a_-\right] + \Delta a_+ = -i C \Bigl(E + \frac{\mu+{\textstyle \frac{1}{4}}}{2m\rho^2}\Bigr) \ , \label{a}
\end{eqnarray} 
where we cancelled $e^{-K(\rho)}$ from all the terms. These equations decouple by taking the sum and difference of $a_+$ and $a_-$
\begin{eqnarray}
&&v_F\frac{d}{d\rho}(a_+ -a_-) - 2 \Delta(a_+ -a_-) = iC \Bigl(2E + \frac{\mu}{m\rho^2}\Bigr) \ , \nonumber \\
&&v_F\frac{d}{d\rho}(a_+ +a_-) = -iC \frac{1}{4m\rho^2} \ . \label{a1}
\end{eqnarray}
The solution of these equations is
\begin{eqnarray}
&&(a_+ -a_-) = -iC \int_{\rho}^{\infty} d\rho' \exp[2K(\rho)-2K(\rho')] \nonumber \\
&&\ \ \ \ \ \ \ \ \ \ \times \Bigl(2\frac{E}{v_F} + \frac{\mu}{k_F\rho'^2} \Bigr) \ , \nonumber \\
&&(a_+ +a_-) = iC \frac{1}{4\rho k_F} \ . \label{a2}
\end{eqnarray}
It is convenient to denote ${\tilde \rho} = k_F \rho$, ${\tilde E} = E/(k_F v_F)$ and ${\tilde \Delta} = \Delta/(k_F v_F)$. It is now straightforward to obtain $a_{\pm}$
\begin{eqnarray}
a_{\pm} &=& \frac{iC}{8 {\tilde \rho}} \mp iC \int_{\tilde \rho}^{\infty} dx e^{2K({\tilde \rho})-2K(x)} \Bigl({\tilde E} +\frac{\mu}{2 x^2}\Bigr) \nonumber \\
&=& \frac{iC}{8 {\tilde \rho}} \pm iC \Bigl( {\tilde E}{\tilde \rho} - \frac{\mu}{2{\tilde \rho}}\Bigr) \nonumber \\
&\mp& iC \int_{\tilde \rho}^{\infty} dx e^{2K({\tilde \rho})-2K(x)} {\tilde \Delta}(x) \Bigl(2{\tilde E}x -\frac{\mu}{x}\Bigr) \ , \label{a3}
\end{eqnarray}
where in the last step we integrated by parts. Below we show that the condition that the integral in expression (\ref{a3}) vanishes yields the energies of the bound states.

To first order in the perturbation we have $C(1 + a_{\pm}) \sim C \exp(a_{\pm})$ and we may write 
\begin{equation}
{\hat g}(\rho) = C \left[\begin{array}{c} e^{a_+} \\ -i e^{a_-} \end{array} \right] \exp[-K(\rho)] \ . \label{g1f}
\end{equation}

The final step consists in matching the solution for small $\rho$ and large $\rho$ at a distance $\rho_c$ from the core of the vortex. The condition that this matching is independent of $\rho_c$ determines a unique wave function valid for all $\rho$ and the energy of the bound state. For this purpose we consider an asymptotic expansion for the Bessel and Hankel functions
\begin{eqnarray}
J_{\nu}(z) &=& \sqrt{\frac{2}{\pi z}} \cos\left[z - \frac{\pi \nu}{2} - \frac{\pi}{4} + \frac{\nu^2-{\textstyle \frac{1}{4}}}{2z}\right] , \label{Bessel1} \\
H^{(1)}_{\nu}(z) &=& \sqrt{\frac{2}{\pi z}} \exp\left[i\Bigl(z - \frac{\pi \nu}{2} - \frac{\pi}{4} + \frac{\nu^2-{\textstyle \frac{1}{4}}}{2z}\Bigr)\right] , \label{Hankel}
\end{eqnarray}
which differs slightly from the ones used by Caroli {\it et al.},\cite{CaroliDGM} but is consistent with the table published by the National Institute of Standards and Technology.\cite{NIST} There are three dependencies on $\rho$ in the special functions, namely, (i) The factor $1/\sqrt{\rho}$ in the Bessel/Hankel functions, (ii) the phase factors $\exp(ik_F \rho)$ and $\exp(\pm i E \rho/v_F)$ of the Bessel/Hankel functions and in ${\hat g}$, and (iii) the dependence on $\exp[i(\nu^2-1/4)/(2k_F\rho)]$.  Note that the complex conjugated function for $\rho > \rho_c$ is also a solution, involving the Hankel function of the second kind. Both solutions are needed to complete the matching.

We explicitly work out the matching for the function $f_+^{\mu}$; the matching for $f_-^{\mu}$ follows similarly. For $\rho < \rho_c$ we have 
\begin{widetext}
\begin{equation}
f_+^{\mu}(\rho_c) = A_+ J_{\mu-1/2}[(k_F \pm q)\rho] = A_+ \sqrt{\frac{2}{\pi (1+{\tilde E}){\tilde \rho}_c}} \cos\left[(1+{\tilde E}){\tilde \rho}_c - \frac{\pi (\mu -{\textstyle \frac{1}{2}})}{2} - \frac{\pi}{4} + \frac{(\mu-{\textstyle \frac{1}{2}})^2-{\textstyle \frac{1}{4}}}{2(1+{\tilde E}){\tilde \rho}_c}\right] \ , \label{smallxx}
\end{equation}
while for $\rho > \rho_c$ we obtain
\begin{equation}
f_+^{\mu}(\rho_c) = [C H^{(1)}_{\mu}({\tilde \rho}_c) g_1({\tilde \rho}_c) + c.c.] =  \frac{1}{2}\sqrt{\frac{2}{\pi{\tilde \rho}_c}} \left\{\exp\left[i\left(\gamma + {\tilde E}{\tilde \rho}_c - \frac{\mu - {\textstyle \frac{1}{4}}}{2{\tilde \rho}_c} +{\tilde \rho}_c - \frac{\pi\mu}{2}- \frac{\pi}{4} + \frac{\mu^2-{\textstyle \frac{1}{4}}}{2{\tilde \rho}_c}\right)\right] + c.c. \right\} e^{-K({\tilde \rho}_c)} \ . \label{largexx}
\end{equation}
\end{widetext}
In Eq. (\ref{largexx}) the phase $\gamma$ arises from $C={\textstyle \frac{1}{2}} e^{i\gamma}$, the next two terms in the exponential are due to $a_+$ and the remainder is consequence of the Hankel function. In Eq. (\ref{smallxx}) $A_+$ is chosen to be $\sqrt{1 + {\tilde E}}$ and since $|{\tilde E}| \ll 1$ we may neglect ${\tilde E}$ in the denominator of the last term. Comparing now the expressions (\ref{smallxx}) and (\ref{largexx}) we find that they are identical for $\gamma = \pi/4$, except for the factor $e^{-K(\rho_c)}$ for the $\rho > \rho_c$ solution. The latter is allowed since for $\rho < \rho_c$ we had neglected $\Delta(\rho)$. Hence, there is a large range for $\rho_c$ where the matching of the solutions is satisfied.

The above hinges on the assumption that the integral in Eq. (\ref{a3}) vanishes, i.e.
\begin{equation}
\int_{{\tilde \rho}_c}^{\infty} dx e^{-2K(x)} 2 {\tilde E} {\tilde \Delta}(x) x = \int_{{\tilde \rho}_c}^{\infty} dx e^{-2K(x)}\frac{\mu {\tilde \Delta}(x)}{x} \ . \label{energy}
\end{equation}
In this expression we may let ${\tilde \rho}_c \to 0$ and integrate the lhs by parts so that
\begin{equation}
E_{\mu} = \mu \int_0^{\infty} dx e^{-2K(x)} \frac{\Delta(x)}{x} \Big/ \int_0^{\infty} dx e^{-2K(x)} \ . \label{energy1}
\end{equation}
Since the main contribution to the integrals is for $\rho \ll \xi$ ($\Delta \approx {\tilde \rho} \Delta_{\infty}^2/E_F$) we obtain $E_{\mu} \approx \mu (\Delta^2_{\infty}/E_F)$, with $\mu = \pm{\textstyle \frac{1}{2}}, \pm{\textstyle \frac{3}{2}}, \pm{\textstyle \frac{5}{2}}, \cdots$. As expected, the excitations are gapped from the Fermi level by a small energy gap of order $\Delta^2_{\infty}/(2E_F)$  (see Ref. [\onlinecite{CaroliDGM}]).

\section{Bound states in a vortex of a 2D Dirac Hamiltonian}

\subsection{Model}

In this Section we consider the 2D Dirac model with $s$-wave superconductivity induced via proximity.\cite{FuKane1,Suzuki,Nori} The electron gas corresponds to the surface states of a topological insulator. The strong spin-orbit interaction couples the spin parallel to the momentum. As before we consider an isolated vortex, assuming a field perpendicular to the plane with $H \ll H_{c2}$ and slightly larger than $H_{c1}$. We apply the same method as used in Section II.

The wave function is a 4-component spinor, $\Psi({\bf r}) = [\psi_{\uparrow}({\bf r}) \ \psi_{\downarrow}({\bf r}) \ \psi^{\dagger}_{\uparrow}({\bf r}) \ \psi^{\dagger}_{\downarrow}({\bf r})]^T$, and the Hamiltonian is ${\cal H} = {\textstyle \frac{1}{2}} \int d^2r \Psi^{\dagger}({\bf r}) {\check {\cal H}}_B^{\parallel}({\bf r}) \Psi({\bf r})$, where
\begin{equation}
{\check {\cal H}}_B^{\parallel}({\bf r}) = \left[ \begin{array}{cc} {\hat h}({\bf r}) & {\hat \Delta({\bf r})} \\
- {\hat \Delta^*({\bf r})} & - {\hat h}^*({\bf r}) \end{array} \right]  \label{HB} 
\end{equation}
and  
\begin{eqnarray}
&&{\hat h}({\bf r}) = v_F {\boldsymbol{\hat \sigma}} \cdot \Bigl({\bf p} - \frac{e}{c} {\bf A} \Bigr) - E_F \ \ , \label{h} \\
&&{\hat \Delta}({\bf r}) = \Delta({\bf r}) i {\hat \sigma}_y \ \ . \label{Delta}
\end{eqnarray}
We adopt polar coordinates, $(\rho,\theta)$, and write $\Delta({\bf r}) = \Delta(\rho) e^{-i\theta}$, i.e. only one flux quantum is contained in the vortex. Using the same arguments as in Refs. [\onlinecite{CaroliDGM}] and [\onlinecite{CaroliM}] we disregard the vector potential in Eq. (\ref{h}). In polar coordinates ${\hat h}({\bf r})$ can be written as
\begin{equation}
{\hat h}(\rho,\theta) = \left[ \begin{array}{cc} -E_F & -i v_F e^{-i\theta} \Bigl(\frac{\partial}{\partial \rho} - \frac{i \ \partial}{\rho \partial \theta} \Bigr) \\ 
-i v_F e^{i\theta} \Bigl(\frac{\partial}{\partial \rho} + \frac{i \ \partial}{\rho \partial \theta} \Bigr) & -E_F \end{array} \right] , \label{h1}
\end{equation}
and the field operators expanded as $\Psi(\rho,\theta) = (2\pi)^{-1/2}  \sum_{\mu} \Psi_{\mu}(\rho) e^{i \mu \theta}$, where $\mu$ is an integer to have a single-valued wave function. As in Sect. II the $\theta$-phase of $\Delta({\bf r})$ can be eliminated via a gauge transformation, yielding a $\theta$ dependence of the components of the spinor $\Psi_{\mu}$ of
\begin{equation}
f_j^{\mu} \exp[-i\theta {\hat \tau}_z (1 + {\hat \sigma}_z)/2 +i \mu \theta] \ \ , \ \ j=1,\cdots,4 \ \ , \label{spinor} 
\end{equation}
where $f_j^{\mu}$ is the amplitude of the component $j$. Applying the spinor to ${\hat h}(\rho,\theta)$ we obtain
\begin{widetext}
\begin{equation}
{\hat h}(\rho,\theta) = \left[ \begin{array}{cc} -E_F & -i v_F e^{-i\theta} \Bigl(\frac{\partial}{\partial \rho} + \frac{\mu}{\rho} \Bigr) \\ 
-i v_F e^{i\theta} \Bigl(\frac{\partial}{\partial \rho} - \frac{\mu-1}{\rho}\Bigr) & -E_F \end{array} \right] \ . \label{h2}
\end{equation}

The first order differential equations satisfied by $f^{\mu}_j$ are 
\begin{eqnarray}
&&-iv_F \left(\frac{\partial}{\partial \rho} - \frac{\mu-1}{\rho}\right) f_1^{\mu}(\rho) - \Delta(\rho) f_3^{\mu}(\rho) -(E +E_F) f_2^{\mu}(\rho) = 0 \ \ , \label{f1} \\
&&-iv_F \left(\frac{\partial}{\partial \rho} + \frac{\mu}{\rho}\right) f_2^{\mu}(\rho) + \Delta(\rho) f_4^{\mu}(\rho) -(E +E_F) f_1^{\mu}(\rho) = 0 \ \ , \label{f2} \\
&&-iv_F \left(\frac{\partial}{\partial \rho} + \frac{\mu+1}{\rho}\right) f_3^{\mu}(\rho) + \Delta(\rho) f_1^{\mu}(\rho) -(E -E_F) f_4^{\mu}(\rho) = 0 \ \ . \label{f3} \\
&&-iv_F \left(\frac{\partial}{\partial \rho} - \frac{\mu}{\rho}\right) f_4^{\mu}(\rho) - \Delta(\rho) f_2^{\mu}(\rho) -(E -E_F) f_3^{\mu}(\rho) = 0 \ \ , \label{f4} 
\end{eqnarray}
\end{widetext}
These equations are analogous to those in Ref. [\onlinecite{Nori}].

\subsection{Majorana state}

The zero-energy Majorana bound state is obtained from Eqs. (\ref{f1}-\ref{f4}) for $E = \mu = 0$. The structure of the equations leads to the solution\cite{Nori}
\begin{eqnarray}
&&f_1^M({\bf r}) = C J_1({\tilde \rho}) e^{-K({\tilde \rho})} e^{-i\theta} \ , \nonumber\\
&&f_2^M({\bf r}) = -i C J_0({\tilde \rho}) e^{-K({\tilde \rho})} \ , \nonumber \\
&&f_3^M({\bf r}) = i C J_1({\tilde \rho}) e^{-K({\tilde \rho})} e^{i\theta} \ , \nonumber \\
&&f_4^M({\bf r}) = - C J_0({\tilde \rho}) e^{-K({\tilde \rho})} \ , \label{Majorana}
\end{eqnarray}
where $C$ is a normalization constant and $K({\tilde \rho}) = \int_0^{\tilde \rho} dx \Delta(x)/E_F$. The corresponding wave function is
\begin{eqnarray}
{\hat \Psi}_{\text M} &=& C \int d^2r e^{-K({\tilde \rho})} \Bigl[J_1({\tilde \rho}) e^{-i\theta} \psi_{\uparrow}({\bf r}) - i J_0({\tilde \rho}) \psi{\downarrow}({\bf r}) \nonumber \\
&& + i J_1({\tilde \rho}) e^{i\theta} \psi^{\dagger}_{\uparrow}({\bf r}) - J_0({\tilde \rho}) \psi^{\dagger}_{\downarrow}({\bf r}) \Bigr] \ \ . \label{Mwfunction}
\end{eqnarray}
It is easily verified that ${\hat \Psi}_{\text M}= - i {\hat \Psi}^{\dagger}_{\text M}$ and hence the state is a Majorana fermion. The counterpart to this Majorana fermion is placed in the plane at large $\rho$ far away from the axis of the vortex and hence not a solution of this problem.\cite{Nori2}

\subsection{Solution for $\rho < \rho_c$}

It is convenient to convert the first order differential equations, (\ref{f1}-\ref{f4}), into second order ones. From Eq. (\ref{f1}) we can express $f^{\mu}_2(\rho)$ and insert it into Eq. (\ref{f2}). Similar substitutions can be done for the remaining equations.  Defining $q_p =(E_F+E)/v_F$ and $q_h =(E_F-E)/v_F$ (for particles and holes, respectively) we obtain
\begin{widetext}
\begin{eqnarray}
&&\left[\frac{{\partial}^2}{\partial \rho^2} + \frac{1}{\rho}\frac{\partial}{\partial \rho} - \frac{(\mu-1)^2}{\rho^2} + q_p^2 \right] f^{\mu}_1 = \frac{q_p}{v_F} \Delta(\rho) f_4^{\mu} + \frac{i}{v_F} \left(\frac{\partial}{\partial \rho} + \frac{\mu}{\rho}\right) \Delta(\rho) f_3^{\mu} \ \ , \label{f1a} \\
&&\left[\frac{{\partial}^2}{\partial \rho^2} + \frac{1}{\rho}\frac{\partial}{\partial \rho} - \frac{\mu^2}{\rho^2} + q_p^2 \right] f^{\mu}_2 = -\frac{q_p}{v_F} \Delta(\rho) f_3^{\mu} - \frac{i}{v_F} \left(\frac{\partial}{\partial \rho} - \frac{\mu-1}{\rho}\right) \Delta(\rho) f_4^{\mu} \ \ , \label{f2a} \\
&&\left[\frac{{\partial}^2}{\partial \rho^2} + \frac{1}{\rho}\frac{\partial}{\partial \rho} - \frac{(\mu+1)^2}{\rho^2} + q_h^2 \right] f^{\mu}_3 = \frac{q_h}{v_F} \Delta(\rho) f_2^{\mu} - \frac{i}{v_F} \left(\frac{\partial}{\partial \rho} - \frac{\mu}{\rho}\right) \Delta(\rho) f_1^{\mu} \ \ , \label{f3a} \\
&&\left[\frac{{\partial}^2}{\partial \rho^2} + \frac{1}{\rho}\frac{\partial}{\partial \rho} - \frac{\mu^2}{\rho^2} + q_h^2 \right] f^{\mu}_4 = -\frac{q_h}{v_F} \Delta(\rho) f_1^{\mu} + \frac{i}{v_F} \left(\frac{\partial}{\partial \rho} + \frac{\mu+1}{\rho}\right) \Delta(\rho) f_2^{\mu} \ \ . \label{f4a} 
\end{eqnarray}
\end{widetext}

Since $\Delta(\rho)$ increases linearly from zero, we may neglect $\Delta(\rho)$ for $\rho < \rho_c$ (as in the previous section). The solutions for $\rho < \rho_c$ are then
\begin{eqnarray}
&&f_1^{\mu}(q_p \rho) = A_1^{\mu} J_{\mu - 1}(q_p \rho) \ \ , \nonumber \\
&&f_2^{\mu}(q_p \rho) = A_2^{\mu} J_{\mu}(q_p \rho) \ \ , \nonumber \\
&&f_3^{\mu}(q_h \rho) = A_3^{\mu} J_{\mu + 1}(q_h \rho) \ \ , \nonumber \\
&&f_4^{\mu}(q_h \rho) = A_4^{\mu} J_{\mu}(q_h \rho) \ \ , \label{Jnu}
\end{eqnarray}  
where $J_{\nu}(z)$ are again Bessel functions. The constants $A_j^{\mu}$ are not all independent. Substituting the solution (\ref{Jnu}) into the first order differential equations we have $A_1^{\mu} = C_1$, $A_2^{\mu} = i C_1$, $A_3^{\mu} = - i C_2$ and $A_4^{\mu} = C_2$. The constants $C_1$ and $C_2$ are independent for $\Delta=0$, but become coupled when $\Delta \neq 0$, namely, $C_1 = \sqrt{1+{\tilde E}}$ and $C_2 = \sqrt{1 - {\tilde E}}$ (see subsection III.E).

\subsection{Solution for $\rho > \rho_c$}

$\Delta$ plays a relevant role for $\rho > \rho_c$. As in Section II, we write the solution as a product of a Hankel function times an envelop function, $f_j({\tilde \rho}) = B_j [H^{(1)}_{\mu}({\tilde \rho}) g_j({\tilde \rho}) + c.c.]$, where the $B_j$ are constants. We postulate $B_1 = {\textstyle \frac{1}{2}}B$, $B_2 = i{\textstyle \frac{1}{2}}B$, $B_3 = -i{\textstyle \frac{1}{2}}B$ and $B_4 = {\textstyle \frac{1}{2}}B$ with $B$ being the normalization constant, which we set equal to one for simplicity. The relative phases of the $B_j$ are the same as in Eq. (\ref{Jnu}). The verification that this choice of $B_j$ is the correct one follows in subsection III.E, where we match the wave function for $\rho<\rho_c$ to $\rho>\rho_c$. We introduce again the dimensionless variables ${\tilde \rho} = k_F \rho$, ${\tilde E} = E/(k_F v_F)$ and ${\tilde \Delta} = \Delta/(k_F v_F)$, and further assume that for $\rho \ll \xi$, $d \Delta({\tilde \rho})/d{\tilde \rho} = \Delta'$, where $\Delta'$ is a constant. 

Next we insert the Ansatz for $f_j({\tilde \rho})$ into Eqs. (\ref{f1a}-\ref{f4a}) and use the differential equation satisfied by the Hankel function. Dividing the equations by $H^{(1)}_{\mu}({\tilde \rho})$ and using Eq. (\ref{asympt}) we arrive at the following four coupled second order differential equations for the functions $g_j({\tilde \rho})$: 
\begin{widetext}
\begin{eqnarray}
&&\frac{d^2 g_1}{d{\tilde \rho}^2} + 2i \frac{d g_1}{d {\tilde \rho}} + \left[{\tilde E}^2 + 2{\tilde E}+\frac{2\mu-1}{{\tilde \rho}^2}\right] g_1 = {\tilde \Delta}(1+{\tilde E}) g_4 + {\tilde \Delta}\frac{d g_3}{d {\tilde \rho}} + \left[\frac{2\mu+1}{2 {\tilde \rho}} + i\right]{\tilde \Delta} g_3 , \label{f1b} \\
&&\frac{d^2 g_2}{d{\tilde \rho}^2} + 2i \frac{d g_2}{d {\tilde \rho}} + \left[{\tilde E}^2 + 2{\tilde E}\right]g_2 = {\tilde \Delta}(1+{\tilde E}) g_3 - {\tilde \Delta}\frac{d g_4}{d {\tilde \rho}} + \left[\frac{2\mu-3}{2 {\tilde \rho}} - i\right]{\tilde \Delta} g_4 , \label{f2b} \\
&&\frac{d^2 g_3}{d{\tilde \rho}^2} + 2i \frac{d g_3}{d {\tilde \rho}} + \left[{\tilde E}^2 - 2{\tilde E}-\frac{2\mu+1}{{\tilde \rho}^2}\right] g_3 = - {\tilde \Delta}(1-{\tilde E}) g_2 + {\tilde \Delta}\frac{d g_1}{d {\tilde \rho}} - \left[\frac{2\mu-1}{2 {\tilde \rho}} - i\right]{\tilde \Delta} g_1 , \label{f3b} \\
&&\frac{d^2 g_4}{d{\tilde \rho}^2} + 2i \frac{d g_4}{d {\tilde \rho}} + \left[{\tilde E}^2 - 2{\tilde E}\right]g_4 = - {\tilde \Delta}(1-{\tilde E}) g_1 - {\tilde \Delta}\frac{d g_2}{d {\tilde \rho}} - \left[\frac{2\mu+3}{2 {\tilde \rho}} + i\right]{\tilde \Delta} g_2 . \label{f4b} 
\end{eqnarray}
\end{widetext}
Here we used Eq. (\ref{asympt}) to simplify derivatives of the Hankel function. Note that the next order correction to Eq. (\ref{asympt}) does not add relevant terms to the order in $1/{\tilde \rho}$ considered here.

As in section II, Eqs. (\ref{f1b}-\ref{f4b}) are solved perturbatively. To zeroth order we have, keeping the dominant terms, 
\begin{equation}
2i\frac{d}{d{\tilde \rho}} \left[\begin{array}{c} g_1^{(0)} \\ g_2^{(0)} \\ g_3^{(0)} \\ g_4^{(0)} \end{array} \right] = {\tilde \Delta} \left[\begin{array}{c} g_4^{(0)} \\  g_3^{(0)} \\ -g_2^{(0)} \\ - g_1^{(0)} \end{array} \right] + i {\tilde \Delta} \left[\begin{array}{c} g_3^{(0)} \\  -g_4^{(0)} \\ g_1^{(0)} \\ - g_2^{(0)} \end{array} \right] \ , \label{lead}
\end{equation}
\begin{widetext}
\noindent
while the remaining terms in Eqs. (\ref{f1b}-\ref{f4b}) will be treated in first order perturbation, $g_j^{(1)}$. The solution of Eq. (\ref{lead}) is
\begin{equation}
g^{(0)}_1({\tilde \rho}) = C e^{-K({\tilde \rho})} \ \ , \ \ g^{(0)}_2({\tilde \rho}) = -i C e^{-K({\tilde \rho})} \ \ , \ \ g^{(0)}_3({\tilde \rho}) = -C e^{-K({\tilde \rho})} \ \ , \ \ g^{(0)}_4({\tilde \rho}) = -i C e^{-K({\tilde \rho})} , \label{g^0}
\end{equation}
where $K({\tilde \rho}) = \int_0^{\tilde \rho} d x {\tilde \Delta}(x)$ is the same function as in section II and $C=e^{i \gamma}$.

The equations for $g_j^{(1)}$ are
\begin{eqnarray}
&&2i\frac{d}{d{\tilde \rho}} \left[\begin{array}{c} g_1^{(1)} \\ g_2^{(1)} \\ g_3^{(1)} \\ g_4^{(1)} \end{array} \right] - {\tilde \Delta} \left[\begin{array}{c} g_4^{(1)} \\  g_3^{(1)} \\ -g_2^{(1)} \\ - g_1^{(1)} \end{array} \right] - i {\tilde \Delta} \left[\begin{array}{c} g_3^{(1)} \\  -g_4^{(1)} \\ g_1^{(1)} \\ - g_2^{(1)} \end{array} \right] = -\frac{d^2}{d{\tilde \rho}^2} \left[\begin{array}{c} g_1^{(0)} \\ g_2^{(0)} \\ g_3^{(0)} \\ g_4^{(0)} \end{array} \right] - \left[\begin{array}{c} ({\tilde E}^2 + 2{\tilde E}+\frac{2\mu-1}{{\tilde \rho}^2}) g_1^{(0)} \\ ({\tilde E}^2 + 2{\tilde E}) g_2^{(0)} \\ ({\tilde E}^2 - 2{\tilde E}-\frac{2\mu+1}{{\tilde \rho}^2}) g_3^{(0)} \\ ({\tilde E}^2 - 2{\tilde E}) g_4^{(0)} \end{array} \right] \nonumber \\
&& \ \ \ \ \ \ \ + {\tilde \Delta}{\tilde E} \left[\begin{array}{c} g_4^{(0)} \\ g_3^{(0)} \\ g_2^{(0)} \\ g_1^{(0)} \end{array} \right] + {\tilde \Delta} \frac{d}{d {\tilde \rho}} \left[\begin{array}{c} g_3^{(0)} \\ -g_4^{(0)} \\ g_1^{(0)} \\ -g_2^{(0)} \end{array} \right] + \frac{\tilde \Delta}{2 {\tilde \rho}} \left[\begin{array}{c} (2\mu+1)g_3^{(0)} \\ (2\mu-3)g_4^{(0)} \\ -(2\mu-1)g_1^{(0)} \\ -(2\mu+3)g_2^{(0)} \end{array} \right], \label{first}        
\end{eqnarray}
We now insert our solutions for $g_j^{(0)}$ into Eq. (\ref{first}), and with the Ansatz $g_1^{(1)} = C a_1 e^{-K}$, $g_2^{(1)} = -iCa_2 e^{-K}$, $g_3^{(1)} = -Ca_3 e^{-K}$ and $g_4^{(1)} = -iCa_4 e^{-K}$, we obtain
\begin{eqnarray}
&&2i\frac{d}{d{\tilde \rho}} \left[\begin{array}{c} a_1 \\ -ia_2 \\ -a_3 \\ -ia_4 \end{array} \right] - 2i {\tilde \Delta} \left[\begin{array}{c} a_1 \\ -ia_2 \\ -a_3 \\ -ia_4 \end{array} \right] - {\tilde \Delta} \left[\begin{array}{c} -ia_4 \\  -a_3 \\ ia_2 \\ - a_1 \end{array} \right] - i {\tilde \Delta} \left[\begin{array}{c} -a_3 \\  ia_4 \\ a_1 \\ ia_2 \end{array} \right] = \left(\frac{\tilde \Delta}{\tilde \rho} - {\tilde \Delta}^2\right) \left[\begin{array}{c} 1 \\ -i \\ -1 \\ -i \end{array} \right] - \left[\begin{array}{c} ({\tilde E}^2 + 2{\tilde E}+\frac{2\mu-1}{{\tilde \rho}^2}) \\ -i ({\tilde E}^2 + 2{\tilde E}) \\ -({\tilde E}^2 - 2{\tilde E}-\frac{2\mu+1}{{\tilde \rho}^2}) \\ -i ({\tilde E}^2 - 2{\tilde E}) \end{array} \right] \nonumber \\
&& \ \ \ \ \ \ \ + {\tilde \Delta}{\tilde E} \left[\begin{array}{c} -i \\ -1 \\ -i \\ 1 \end{array} \right] - {\tilde \Delta}^2 \left[\begin{array}{c} -1 \\ i \\ 1 \\ i \end{array} \right] + \frac{\tilde \Delta}{2 {\tilde \rho}} \left[\begin{array}{c} -(2\mu+1) \\ -i(2\mu-3) \\ -(2\mu-1) \\ i(2\mu+3) \end{array} \right]. \label{second}        
\end{eqnarray}
Since all the terms are proportional to $e^{-K({\tilde \rho})}$, this factor has been cancelled out. After cancellations of terms, the differential equations for $a_j$ are
\begin{equation}
2\frac{d}{d{\tilde \rho}} \left[\begin{array}{c} a_1 \\ -ia_2 \\ -a_3 \\ -ia_4 \end{array} \right] + {\tilde \Delta} \left[\begin{array}{c} (a_3+a_4)-2a_1 \\ (a_3+a_4)-2a_2 \\ (a_1+a_2)-2a_3 \\ (a_1+a_2)-2a_4 \end{array} \right] =  - \left[\begin{array}{c} i({\tilde E}^2 + 2{\tilde E}+\frac{2\mu-1}{{\tilde \rho}^2}) \\ i ({\tilde E}^2 + 2{\tilde E}) \\ i({\tilde E}^2 - 2{\tilde E}-\frac{2\mu+1}{{\tilde \rho}^2}) \\ i ({\tilde E}^2 - 2{\tilde E}) \end{array} \right] + {\tilde \Delta}{\tilde E} \left[\begin{array}{c} -1 \\ -1 \\ 1 \\ 1 \end{array} \right] + \frac{\tilde \Delta}{2{\tilde \rho}} \left[\begin{array}{c} i(2\mu-1) \\ -i(2\mu-1) \\ -i(2\mu+1) \\ i(2\mu+1) \end{array} \right]. \label{third}     
\end{equation}
These equations decouple by taking linear combinations:
\begin{eqnarray}
&&2 \frac{d}{d{\tilde \rho}} (a_1 - a_2) - 2{\tilde \Delta}(a_1 - a_2) = i \frac{2\mu-1}{{\tilde \rho}^2} + i\frac{\tilde \Delta}{\tilde \rho}(2\mu-1) , \nonumber \\ 
&&2 \frac{d}{d{\tilde \rho}} (a_3 - a_4) - 2{\tilde \Delta}(a_3 - a_4) = - i \frac{2\mu+1}{{\tilde \rho}^2} - i\frac{\tilde \Delta}{\tilde \rho}(2\mu+1) , \nonumber \\  
&&\frac{d}{d{\tilde \rho}} (a_1+a_2+ a_3 + a_4)= 2 i {\tilde E}^2 - i \frac{1}{{\tilde \rho}^2} , \nonumber \\  
&&\frac{d}{d{\tilde \rho}} (a_1+a_2- a_3 - a_4)- 2{\tilde \Delta}(a_1+a_2- a_3 - a_4) = 4 i {\tilde E} + i \frac{2\mu}{{\tilde \rho}^2} -2{\tilde \Delta}{\tilde E}, \label{third1-4}        
\end{eqnarray}

The integration of the decoupled differential equations yields 
\begin{eqnarray}
&&a_1 - a_2 = -i (\mu -{\textstyle \frac{1}{2}}) \int_{\tilde \rho}^{\infty} dx \exp[K({\tilde \rho}) - K(x)] \left[ \frac{1}{x^2} + \frac{{\tilde \Delta}(x)}{x} \right] \ , \label{xx1} \\
&&a_3 - a_4 = i (\mu +{\textstyle \frac{1}{2}}) \int_{\tilde \rho}^{\infty} dx \exp[K({\tilde \rho}) - K(x)] \left[ \frac{1}{x^2} + \frac{{\tilde \Delta}(x)}{x} \right] \ , \label{xx2} \\
&&a_1+a_2+a_3+a_4 = 2i {\tilde E}^2 {\tilde \rho} + \frac{i}{\tilde \rho} \ , \label{xx3} \\
&&a_1+a_2-a_3-a_4 = -i \int_{\tilde \rho}^{\infty} dx \exp[2K({\tilde \rho}) - 2K(x)] \left[4{\tilde E} + \frac{2\mu}{x^2} \right] - 2\int_0^{\tilde \rho} dx \exp[2K({\tilde \rho}) - 2K(x)] {\tilde E}{\tilde \Delta}(x) \ . \label{xx4}        
\end{eqnarray}
The first term in Eq. (\ref{xx3}) and the last term in Eq. (\ref{xx4}) are of third order in the small parameters ${\tilde \Delta({\tilde \rho})}$, ${\tilde E}$ and ${\tilde \rho}$ and can be neglected. The remaining two integrals can be simplified by integrating by parts
\begin{eqnarray}
&&\int_{\tilde \rho}^{\infty} dx \exp[2K({\tilde \rho}) - 2K(x)] \left[{\tilde E} + \frac{\mu}{2 x^2} \right] = - {\tilde E}{\tilde \rho} + \frac{\mu}{2 {\tilde \rho}} + \int_{\tilde \rho}^{\infty} dx \exp[2K({\tilde \rho}) - 2K(x)]\Bigl[2 {\tilde \Delta}(x) {\tilde E} x - \frac{\mu {\tilde \Delta}(x)}{x}\Bigr] \ , \nonumber \\ 
&&\int_{\tilde \rho}^{\infty} dx \exp[K({\tilde \rho}) - K(x)] \Bigl[ \frac{1}{x^2} + \frac{{\tilde \Delta}(x)}{x} \Bigr] = \frac{1}{2 {\tilde \rho}} \ . \label{byparts}
\end{eqnarray}
It is straightforward to solve the above equations for the $a_j$:
\begin{eqnarray}
&&a_1 = i\Bigl({\tilde E}{\tilde \rho} - \frac{2\mu-1}{2 {\tilde \rho}}\Bigr) - i\int_{\tilde \rho}^{\infty} dx \exp[2K({\tilde \rho}) - 2K(x)] \ {\tilde \Delta}(x) \left[2 {\tilde E} x - \frac{\mu}{x} \right] \ , \label{a1} \\
&&a_2 = i {\tilde E}{\tilde \rho} - i\int_{\tilde \rho}^{\infty} dx \exp[2K({\tilde \rho}) - 2K(x)] \ {\tilde \Delta}(x) \left[2 {\tilde E} x - \frac{\mu}{x} \right] \ , \label{a2} \\
&&a_3 = -i\Bigl({\tilde E}{\tilde \rho} - \frac{2\mu+1}{2 {\tilde \rho}}\Bigr) + i\int_{\tilde \rho}^{\infty} dx \exp[2K({\tilde \rho}) - 2K(x)] \ {\tilde \Delta}(x) \left[2 {\tilde E} x - \frac{\mu}{x} \right] \ , \label{a3} \\
&&a_4 = -i {\tilde E}{\tilde \rho} + i\int_{\tilde \rho}^{\infty} dx \exp[2K({\tilde \rho}) - 2K(x)] \ {\tilde \Delta}(x) \left[2 {\tilde E} x - \frac{\mu}{x} \right] \ . \label{a4} 
\end{eqnarray}
\end{widetext}
The common integral term in Eqs. (\ref{a1})-(\ref{a4}) is zero and defines the bound state energy.

\subsection{Matching of wave functions}

The solutions for ${\tilde \rho} > {\tilde \rho}_c$ are now given by $f^{\mu}_j({\tilde \rho}) = B_j [H^{(1)}_{\mu}({\tilde \rho}) g_j({\tilde \rho}) + c.c.]$, where the coefficients $B_j$ are defined at the beginning of subsection III.D. They have to be matched at ${\tilde \rho}_c$ to $f^{\mu}_j({\tilde \rho})$ given by Eq. (\ref{Jnu}) for ${\tilde \rho} < {\tilde \rho}_c$ using a similar procedure to that of Sect. II and Ref. \onlinecite{CaroliDGM}. The functions $g_j$ were calculated consistently to first order of perturbation and can be written as an exponential, i.e. $g_j({\tilde \rho}) \propto \exp[-K({\tilde \rho}) + a_j({\tilde \rho})]$, which remains correct to first order. To match the wave functions we employ the asymptotic expansions for the Bessel and the Hankel functions shown in Eqs. (\ref{Bessel1}) and (\ref{Hankel}).

As shown in Section II there are three factors in $f_j$ depending on ${\tilde \rho}$: (i) The $1/\sqrt{\tilde \rho}_c$-dependence in the Bessel and Hankel functions, (ii) the phase factors of the form $\exp[i(1 \pm {\tilde E}){\tilde \rho}_c]$, and (iii) the factors $\exp\{i[(\mu\pm1)^2-1/4]/[2{\tilde \rho}_c]\}$. We explicitly work out the matching for the function $f_1^{\mu}$; the other three functions follow similarly. For ${\tilde \rho} < {\tilde \rho}_c$ we have 
\begin{widetext}
\begin{equation}
f_1^{\mu}({\tilde \rho}_c) = C_1 J_{\mu - 1}(q_p \rho_c) = C_1 \sqrt{\frac{2}{\pi (1+{\tilde E}) {\tilde \rho}_c}} \cos\left[(1+{\tilde E}) {\tilde \rho}_c - \frac{\pi (\mu -1)}{2} - \frac{\pi}{4} + \frac{(\mu-1)^2-{\textstyle \frac{1}{4}}}{2(1+{\tilde E}){\tilde \rho}_c}\right] \ , \label{small}
\end{equation}
while for ${\tilde \rho} > {\tilde \rho}_c$ we obtained
\begin{equation}
f_1^{\mu}({\tilde \rho}_c) = B_1 [H^{(1)}_{\mu}({\tilde \rho}_c) g_1({\tilde \rho}_c) + c.c.] =  \frac{1}{2}\sqrt{\frac{2}{\pi{\tilde \rho}_c}} \left\{\exp\left[i\left(\gamma + {\tilde E}{\tilde \rho}_c - \frac{(2\mu -1)}{2{\tilde \rho}_c} +{\tilde \rho}_c - \frac{\pi\mu}{2}- \frac{\pi}{4} + \frac{\mu^2-{\textstyle \frac{1}{4}}}{2{\tilde \rho}_c}\right)\right] + c.c. \right\} e^{-K({\tilde \rho}_c)} \ . \label{large}
\end{equation}
\end{widetext}
In Eq. (\ref{large}) the phase $\gamma$ arises from $C$, Eq. (\ref{g^0}), the next two terms in the exponential are due to $a_1$ and the remainder is consequence of the Hankel function.

From comparing these two expressions it follows that $C_1=\sqrt{1+{\tilde E}}$ and $\gamma = \pi/2$. As in Ref. [\onlinecite{CaroliDGM}] and Section II we neglect (1) ${\tilde E}$ in the denominator of the last term in Eq. (\ref{small}), because ${\tilde E} \ll 1$, and (2) the factor $e^{-K({\tilde \rho}_c)}$ in Eq. (\ref{large}). The two expressions are then equivalent and the matching is satisfied for a large interval of ${\tilde \rho}_c$.

The bound state energies are then determined by the integral term in Eqs. (\ref{a1})-(\ref{a4}), which has to vanish. At this point we can take ${\tilde \rho}_c \to 0$ in the lower integration limit and integrate the first term by parts:
\begin{equation}
E_{\mu} = \mu \int_0^{\infty} dx e^{-2K(x)} \frac{\Delta(x)}{x} \Big/ \int_0^{\infty} dx  e^{-2K(x)} \ . \label{bound}
\end{equation}
Since the main contribution to the integrals is for $\rho \ll \xi$, where $\Delta(x)$ is linear in $x$, we arrive at $E_{\mu} \approx \mu \Delta'/k_F$, where $\Delta' =d\Delta/d\rho \approx k_F\Delta_{\infty}/\xi \approx k_F\Delta_{\infty}^2/E_F$ and hence
\begin{equation}
E_{\mu} \approx  \mu \frac{\Delta_{\infty}^2}{E_F}  \ . \label{bound1}
\end{equation}
The above holds for all integer values of $\mu$.  

Approximate expressions for the amplitudes are given by 
\begin{eqnarray}
&&f_1^{\mu}({\bf r}) = J_{\mu - 1}({\tilde \rho}) e^{-K({\tilde \rho})} e^{-i\theta} \ , \nonumber\\
&&f_2^{\mu}({\bf r}) = i J_{\mu}({\tilde \rho}) e^{-K({\tilde \rho})}  \ , \nonumber \\
&&f_3^{\mu}({\bf r}) = -i J_{\mu + 1}({\tilde \rho}) e^{-K({\tilde \rho})} e^{i\theta} \ , \nonumber \\
&&f_4^{\mu}({\bf r}) = J_{\mu}({\tilde \rho}) e^{-K({\tilde \rho})}  \ , \label{amplitudes}
\end{eqnarray}
and the energy wave function with energy $E=E_{\mu}$ is then 
\begin{widetext}
\begin{equation}
{\hat \Psi}_E = \int d^2r e^{-K({\tilde \rho})} \Bigl[J_{\mu - 1}({\tilde \rho}) e^{-i\theta} \psi_{\uparrow}({\bf r}) + i J_{\mu}({\tilde \rho}) \psi{\downarrow}({\bf r}) 
- i J_{\mu + 1}({\tilde \rho}) e^{i\theta} \psi^{\dagger}_{\uparrow}({\bf r}) + J_{\mu}({\tilde \rho}) \psi^{\dagger}_{\downarrow}({\bf r}) \Bigr] \ \ . \label{wfunction}
\end{equation}
\end{widetext}
For $\mu \ne 0$ the wave function corresponds to a fermion operator with $E_{\mu} \neq 0$, while for $\mu = 0$ we have the Majorana state wave function consistent with Eq. (\ref{Mwfunction}) up to an overall minus sign by noticing $J_{-1}(\tilde{\rho})=-J_{1}(\tilde{\rho})$. 

\section{Case of perpendicular spin and momentum locking}

\subsection{Model}

In this section we consider a Hamiltonian with the spin-orbit coupling given by the Rashba interaction\cite{Loss,Bobkova}
\begin{equation}
{\hat h}({\bf r}) = v_F \boldsymbol{\hat \sigma} \cdot \Bigl[\Bigl({\bf p} - \frac{e}{c} {\bf A} \Bigr)\times {\bf e}_z\Bigr] - E_F \ , \label{h1} 
\end{equation}
where ${\bf e}_z$ is the normal vector to the plane. The 4-component wave function, the structure of ${\check {\cal H}}_B^{\perp}$ and the order parameter, Eqs. (\ref{HB}) and (\ref{Delta}), remain unchanged. In polar coordinates $\Delta({\bf r}) = \Delta(\rho) e^{-i\theta}$ and 
\begin{equation}
{\hat h}(\rho,\theta) = \left[ \begin{array}{cc} -E_F & -i v_F e^{-i\theta} \Bigl(i\frac{\partial}{\partial \rho} + \frac{1}{\rho} \frac{\partial}{\partial \theta} \Bigr) \\ 
i v_F e^{i\theta} \Bigl(i\frac{\partial}{\partial \rho} - \frac{1}{\rho} \frac{\partial}{\partial \theta} \Bigr) & -E_F \end{array} \right] . \label{h1a}
\end{equation}
Here we neglected the vector potential following the same arguments as in Refs. [\onlinecite{CaroliDGM}] and [\onlinecite{CaroliM}] and previous sections.  As in Sect. III we expand the field operators as $\Psi(\rho,\theta) = (2\pi)^{-1/2}  \sum_{\mu} \Psi_{\mu}(\rho) e^{i \mu \theta}$, where $\mu$ is an integer to have a single-valued wave function,  
and we eliminate $\theta$-phase of $\Delta({\bf r})$ via a gauge transformation. The $\theta$ dependence of the components of the spinor $\Psi_{\mu}$ is again given by Eq. (\ref{spinor}). Applying the spinor $\Psi_{\mu}(\rho) e^{i \mu \theta}$ to Eq. (\ref{h1a}) we obtain
\begin{equation}
{\hat h}_{\mu}(\rho) = \left[ \begin{array}{cc} -E_F & v_F \Bigl(\frac{\partial}{\partial \rho} + \frac{\mu}{\rho} \Bigr) \\ 
-v_F \Bigl(\frac{\partial}{\partial \rho} - \frac{\mu-1}{\rho}\Bigr) & -E_F \end{array} \right] \ . \label{h2}
\end{equation}
As in Sect. III we denote with $f_j^{\mu}$ the amplitude of the component $j$ of the spinor. The equations of motion for the amplitudes are similar to Eqs. (\ref{f2}-\ref{f3}), except for factors $i$ and signs,
\begin{widetext}
\begin{eqnarray}
&&v_F \left(\frac{\partial}{\partial \rho} - \frac{\mu-1}{\rho}\right) f_1^{\mu}(\rho) + \Delta(\rho) f_3^{\mu}(\rho) +(E +E_F) f_2^{\mu}(\rho) = 0 \ \ , \label{f1c} \\
&&v_F \left(\frac{\partial}{\partial \rho} + \frac{\mu}{\rho}\right) f_2^{\mu}(\rho) + \Delta(\rho) f_4^{\mu}(\rho) -(E +E_F) f_1^{\mu}(\rho) = 0 \ \ , \label{f2c} \\
&&v_F \left(\frac{\partial}{\partial \rho} + \frac{\mu+1}{\rho}\right) f_3^{\mu}(\rho) + \Delta(\rho) f_1^{\mu}(\rho) -(E -E_F) f_4^{\mu}(\rho) = 0 \ \ , \label{f3c} \\
&&v_F \left(\frac{\partial}{\partial \rho} - \frac{\mu}{\rho}\right) f_4^{\mu}(\rho) + \Delta(\rho) f_2^{\mu}(\rho) +(E -E_F) f_3^{\mu}(\rho) = 0 \ \ . \label{f4c} 
\end{eqnarray}
\end{widetext}

\subsection{Majorana state}

The zero-energy Majorana bound state is obtained from Eqs. (\ref{f1c}-\ref{f4c}) for $E = \mu = 0$. The structure of the equations leads to the solution
\begin{eqnarray}
&&f_1^M({\bf r}) = C J_1({\tilde \rho}) e^{-K({\tilde \rho})} e^{-i\theta} \ , \nonumber\\
&&f_2^M({\bf r}) = - C J_0({\tilde \rho}) e^{-K({\tilde \rho})} \ , \nonumber \\
&&f_3^M({\bf r}) = C J_1({\tilde \rho}) e^{-K({\tilde \rho})} e^{i\theta} \ , \nonumber \\
&&f_4^M({\bf r}) = - C J_0({\tilde \rho}) e^{-K({\tilde \rho})} \ , \label{Majorana}
\end{eqnarray}
where $C$ is a normalization constant and $K({\tilde \rho}) = \int_0^{\tilde \rho} dx \Delta(x)/E_F$. The corresponding wave function is
\begin{eqnarray}
{\hat \Psi}_{\text M} &=& C \int d^2r e^{-K({\tilde \rho})} \Bigl[J_1({\tilde \rho}) e^{-i\theta} \psi_{\uparrow}({\bf r}) - J_0({\tilde \rho}) \psi{\downarrow}({\bf r}) \nonumber \\
&& + J_1({\tilde \rho}) e^{i\theta} \psi^{\dagger}_{\uparrow}({\bf r}) - J_0({\tilde \rho}) \psi^{\dagger}_{\downarrow}({\bf r}) \Bigr] \ \ . \label{Mwfunction1}
\end{eqnarray}
It is easily verified that ${\hat \Psi}_{\text M} = {\hat \Psi}^{\dagger}_{\text M}$ and hence the state is a Majorana fermion. The counterpart to this Majorana fermion is placed in the plane far away from the axis of the vortex (large $\rho$ and hence not a solution of this problem.\cite{Nori2}

\subsection{Solution for $\rho < \rho_c$}

With similar substitutions as in Sect. III we convert the first order differential equations, (\ref{f1c})-(\ref{f4c}), into second order ones
\begin{widetext}
\begin{eqnarray}
&&\left[\frac{{\partial}^2}{\partial \rho^2} + \frac{1}{\rho}\frac{\partial}{\partial \rho} - \frac{(\mu-1)^2}{\rho^2} + q_p^2 \right] f^{\mu}_1 = \frac{q_p}{v_F} \Delta(\rho) f_4^{\mu} - \left(\frac{\partial}{\partial \rho} + \frac{\mu}{\rho}\right) \frac{\Delta(\rho)}{v_F} f_3^{\mu} \ \ , \label{f1d} \\
&&\left[\frac{{\partial}^2}{\partial \rho^2} + \frac{1}{\rho}\frac{\partial}{\partial \rho} - \frac{\mu^2}{\rho^2} + q_p^2 \right] f^{\mu}_2 = -\frac{q_p}{v_F} \Delta(\rho) f_3^{\mu} - \left(\frac{\partial}{\partial \rho} - \frac{\mu-1}{\rho}\right) \frac{\Delta(\rho)}{v_F} f_4^{\mu} \ \ , \label{f2d} \\
&&\left[\frac{{\partial}^2}{\partial \rho^2} + \frac{1}{\rho}\frac{\partial}{\partial \rho} - \frac{(\mu+1)^2}{\rho^2} + q_h^2 \right] f^{\mu}_3 = \frac{q_h}{v_F} \Delta(\rho) f_2^{\mu} - \left(\frac{\partial}{\partial \rho} - \frac{\mu}{\rho}\right) \frac{\Delta(\rho)}{v_F} f_1^{\mu} \ \ , \label{f3d} \\
&&\left[\frac{{\partial}^2}{\partial \rho^2} + \frac{1}{\rho}\frac{\partial}{\partial \rho} - \frac{\mu^2}{\rho^2} + q_h^2 \right] f^{\mu}_4 = -\frac{q_h}{v_F} \Delta(\rho) f_1^{\mu} - \left(\frac{\partial}{\partial \rho} + \frac{\mu+1}{\rho}\right) \frac{\Delta(\rho)}{v_F} f_2^{\mu} \ \ . \label{f4d} 
\end{eqnarray}
\end{widetext}
As previously, since $\Delta(\rho)$ increases linearly from zero, we may neglect $\Delta(\rho)$ for $\rho < \rho_c$.  In terms of Bessel functions the solutions for $\rho < \rho_c$ are then identical to Eq. (\ref{Jnu}), but with the constants $A_1 = A_2$ and $A_3 = -A_4$.  The constants $A_1$ and $A_4$ are independent for $\Delta=0$, but become coupled when $\Delta \neq 0$, namely, $A_1 = \sqrt{1+{\tilde E}}$ and $A_4 = \sqrt{1-{\tilde E}}$.

\subsection{Solution for $\rho > \rho_c$}

For $\rho > \rho_c$, we again write the solution as a product of a Hankel function times an envelop function, $f_j({\tilde \rho}) = B_j [H^{(1)}_{\mu}({\tilde \rho}) g_j({\tilde \rho}) + c.c.]$, and as previously we denote ${\tilde \rho} = k_F \rho$. We assume $B_1 = B_2 = - B_3 = B_4 = \frac{1}{2} B$, where $B$ is the normalization constant equated to one in the following. The matching of the wave function for $\rho < \rho_c$ and $\rho > \rho_c$ in subsection IV.E shows that this assumption is consistent. Inserting the Ansatz into Eqs. (\ref{f1d}-\ref{f4d}) and using the differential equation satisfied by the Hankel function we obtain second order differential equations for the functions $g_j({\tilde \rho})$: 

\noindent
\begin{widetext}
\begin{eqnarray}
&&\frac{d^2 g_1}{d{\tilde \rho}^2} + 2i \frac{d g_1}{d {\tilde \rho}} + \left[{\tilde E}^2 + 2{\tilde E}+\frac{2\mu-1}{{\tilde \rho}^2}\right] g_1 = {\tilde \Delta}(1+{\tilde E}) g_4 + {\tilde \Delta}\frac{d g_3}{d {\tilde \rho}} + \left[\frac{2\mu+1}{2 {\tilde \rho}} + i\right]{\tilde \Delta} g_3 , \label{f1e} \\
&&\frac{d^2 g_2}{d{\tilde \rho}^2} + 2i \frac{d g_2}{d {\tilde \rho}} + \left[{\tilde E}^2 + 2{\tilde E}\right]g_2 = {\tilde \Delta}(1+{\tilde E}) g_3 - {\tilde \Delta}\frac{d g_4}{d {\tilde \rho}} + \left[\frac{2\mu-3}{2 {\tilde \rho}} - i\right]{\tilde \Delta} g_4 , \label{f2e} \\
&&\frac{d^2 g_3}{d{\tilde \rho}^2} + 2i \frac{d g_3}{d {\tilde \rho}} + \left[{\tilde E}^2 - 2{\tilde E}-\frac{2\mu+1}{{\tilde \rho}^2}\right] g_3 = - {\tilde \Delta}(1-{\tilde E}) g_2 + {\tilde \Delta}\frac{d g_1}{d {\tilde \rho}} - \left[\frac{2\mu-1}{2 {\tilde \rho}} - i\right]{\tilde \Delta} g_1 , \label{f3e} \\
&&\frac{d^2 g_4}{d{\tilde \rho}^2} + 2i \frac{d g_4}{d {\tilde \rho}} + \left[{\tilde E}^2 - 2{\tilde E}\right]g_4 = - {\tilde \Delta}(1-{\tilde E}) g_1 - {\tilde \Delta}\frac{d g_2}{d {\tilde \rho}} - \left[\frac{2\mu+3}{2 {\tilde \rho}} + i\right]{\tilde \Delta} g_2 . \label{f4e} 
\end{eqnarray}
Here we used Eq. (\ref{asympt}). Note that the next order correction to Eq. (\ref{asympt}) does not add relevant terms to the order in $1/{\tilde \rho}$ considered here.

As in sections II and III, Eqs. (\ref{f1e}-\ref{f4e}) are solved perturbatively. To zeroth order we have, keeping the dominant terms, 
\begin{equation}
2i\frac{d}{d{\tilde \rho}} \left[\begin{array}{c} g_1^{(0)} \\ g_2^{(0)} \\ g_3^{(0)} \\ g_4^{(0)} \end{array} \right] = {\tilde \Delta} \left[\begin{array}{c} g_4^{(0)} \\  g_3^{(0)} \\ -g_2^{(0)} \\ - g_1^{(0)} \end{array} \right] + i {\tilde \Delta} \left[\begin{array}{c} g_3^{(0)} \\  -g_4^{(0)} \\ g_1^{(0)} \\ - g_2^{(0)} \end{array} \right] \ , \label{lead1}
\end{equation}
while the remaining terms in Eqs. (\ref{f1e}-\ref{f4e}) will be treated in first order perturbation, $g_j^{(1)}$. The solution of Eq. (\ref{lead1}) is
\begin{equation}
g^{(0)}_1({\tilde \rho}) = C e^{-K({\tilde \rho})} \ \ , \ \ g^{(0)}_2({\tilde \rho}) = -i C e^{-K({\tilde \rho})} \ \ , \ \ g^{(0)}_3({\tilde \rho}) = -C e^{-K({\tilde \rho})} \ \ , \ \ g^{(0)}_4({\tilde \rho}) = -i C e^{-K({\tilde \rho})} , \label{B}
\end{equation}
where $C = e^{i\gamma}$ and $\gamma$ is to be determined.

The equations for $g_j^{(1)}$ are
\begin{eqnarray}
&&2i\frac{d}{d{\tilde \rho}} \left[\begin{array}{c} g_1^{(1)} \\ g_2^{(1)} \\ g_3^{(1)} \\ g_4^{(1)} \end{array} \right] - {\tilde \Delta} \left[\begin{array}{c} g_4^{(1)} \\  g_3^{(1)} \\ -g_2^{(1)} \\ - g_1^{(1)} \end{array} \right] - i {\tilde \Delta} \left[\begin{array}{c} g_3^{(1)} \\  -g_4^{(1)} \\ g_1^{(1)} \\ - g_2^{(1)} \end{array} \right] = -\frac{d^2}{d{\tilde \rho}^2} \left[\begin{array}{c} g_1^{(0)} \\ g_2^{(0)} \\ g_3^{(0)} \\ g_4^{(0)} \end{array} \right] - \left[\begin{array}{c} ({\tilde E}^2 + 2{\tilde E}+\frac{2\mu-1}{{\tilde \rho}^2}) g_1^{(0)} \\ ({\tilde E}^2 + 2{\tilde E}) g_2^{(0)} \\ ({\tilde E}^2 - 2{\tilde E}-\frac{2\mu+1}{{\tilde \rho}^2}) g_3^{(0)} \\ ({\tilde E}^2 - 2{\tilde E}) g_4^{(0)} \end{array} \right] \nonumber \\
&& \ \ \ \ \ \ \ + {\tilde \Delta}{\tilde E} \left[\begin{array}{c} g_4^{(0)} \\ g_3^{(0)} \\ g_2^{(0)} \\ g_1^{(0)} \end{array} \right] + {\tilde \Delta} \frac{d}{d {\tilde \rho}} \left[\begin{array}{c} g_3^{(0)} \\ -g_4^{(0)} \\ g_1^{(0)} \\ -g_2^{(0)} \end{array} \right] + \frac{\tilde \Delta}{2 {\tilde \rho}} \left[\begin{array}{c} (2\mu+1)g_3^{(0)} \\ (2\mu-3)g_4^{(0)} \\ -(2\mu-1)g_1^{(0)} \\ -(2\mu+3)g_2^{(0)} \end{array} \right]. \label{first1}        
\end{eqnarray}
Inserting our solutions for $g_j^{(0)}$ into Eq. (\ref{first1}), and with the Ansatz $g_1^{(1)} = C a_1 e^{-K}$, $g_2^{(1)} = -iCa_2 e^{-K}$, $g_3^{(1)} = -Ca_3 e^{-K}$ and $g_4^{(1)} = -iCa_4 e^{-K}$, we obtain
\begin{eqnarray}
&&2i\frac{d}{d{\tilde \rho}} \left[\begin{array}{c} a_1 \\ -ia_2 \\ -a_3 \\ -ia_4 \end{array} \right] - 2i {\tilde \Delta} \left[\begin{array}{c} a_1 \\ -ia_2 \\ -a_3 \\ -ia_4 \end{array} \right] - {\tilde \Delta} \left[\begin{array}{c} -ia_4 \\  -a_3 \\ ia_2 \\ - a_1 \end{array} \right] - i {\tilde \Delta} \left[\begin{array}{c} -a_3 \\  ia_4 \\ a_1 \\ ia_2 \end{array} \right] = \left(\frac{\tilde \Delta}{\tilde \rho} - {\tilde \Delta}^2\right) \left[\begin{array}{c} 1 \\ -i \\ -1 \\ -i \end{array} \right] - \left[\begin{array}{c} ({\tilde E}^2 + 2{\tilde E}+\frac{2\mu-1}{{\tilde \rho}^2}) \\ -i ({\tilde E}^2 + 2{\tilde E}) \\ -({\tilde E}^2 - 2{\tilde E}-\frac{2\mu+1}{{\tilde \rho}^2}) \\ -i ({\tilde E}^2 - 2{\tilde E}) \end{array} \right] \nonumber \\
&& \ \ \ \ \ \ \ + {\tilde \Delta}{\tilde E} \left[\begin{array}{c} -i \\ -1 \\ -i \\ 1 \end{array} \right] - {\tilde \Delta}^2 \left[\begin{array}{c} -1 \\ i \\ 1 \\ i \end{array} \right] + \frac{\tilde \Delta}{2 {\tilde \rho}} \left[\begin{array}{c} -(2\mu+1) \\ -i(2\mu-3) \\ -(2\mu-1) \\ i(2\mu+3) \end{array} \right]. \label{second1}        
\end{eqnarray}
Since all the terms are proportional to $e^{-K({\tilde \rho})}$, this factor has been cancelled out. After cancellations of ${\tilde \Delta}^2$-terms, the differential equations for $a_j$ are
\begin{equation}
2\frac{d}{d{\tilde \rho}} \left[\begin{array}{c} a_1 \\ -ia_2 \\ -a_3 \\ -ia_4 \end{array} \right] + {\tilde \Delta} \left[\begin{array}{c} (a_3+a_4)-2a_1 \\ (a_3+a_4)-2a_2 \\ (a_1+a_2)-2a_3 \\ (a_1+a_2)-2a_4 \end{array} \right] =  - \left[\begin{array}{c} i({\tilde E}^2 + 2{\tilde E}+\frac{2\mu-1}{{\tilde \rho}^2}) \\ i ({\tilde E}^2 + 2{\tilde E}) \\ i({\tilde E}^2 - 2{\tilde E}-\frac{2\mu+1}{{\tilde \rho}^2}) \\ i ({\tilde E}^2 - 2{\tilde E}) \end{array} \right] + {\tilde \Delta}{\tilde E} \left[\begin{array}{c} -1 \\ -1 \\ 1 \\ 1 \end{array} \right] + \frac{\tilde \Delta}{2{\tilde \rho}} \left[\begin{array}{c} i(2\mu-1) \\ -i(2\mu-1) \\ -i(2\mu+1) \\ i(2\mu+1) \end{array} \right]. \label{third1}     
\end{equation}

These equations decouple by taking linear combinations:
\begin{eqnarray}
&&2 \frac{d}{d{\tilde \rho}} (a_1 - a_2) - 2{\tilde \Delta}(a_1 - a_2) = i \frac{2\mu-1}{{\tilde \rho}^2} + i\frac{\tilde \Delta}{\tilde \rho}(2\mu-1) , \nonumber \\ 
&&2 \frac{d}{d{\tilde \rho}} (a_3 - a_4) - 2{\tilde \Delta}(a_3 - a_4) = - i \frac{2\mu+1}{{\tilde \rho}^2} - i\frac{\tilde \Delta}{\tilde \rho}(2\mu+1) , \nonumber \\  
&&\frac{d}{d{\tilde \rho}} (a_1+a_2+ a_3 + a_4)= 2 i {\tilde E}^2 - i \frac{1}{{\tilde \rho}^2} , \nonumber \\  
&&\frac{d}{d{\tilde \rho}} (a_1+a_2- a_3 - a_4)- 2{\tilde \Delta}(a_1+a_2- a_3 - a_4) = 4 i {\tilde E} + i \frac{2\mu}{{\tilde \rho}^2} -2{\tilde \Delta}{\tilde E}, \label{third1-4}        
\end{eqnarray}
The integration of the decoupled differential equations yields
\begin{eqnarray}
&&a_1 - a_2 = -i (\mu -{\textstyle \frac{1}{2}}) \int_{\tilde \rho}^{\infty} dx \exp[K({\tilde \rho}) - K(x)] \left[ \frac{1}{x^2} + \frac{{\tilde \Delta}(x)}{x} \right] \ , \label{xy1} \\
&&a_3 - a_4 = i (\mu +{\textstyle \frac{1}{2}}) \int_{\tilde \rho}^{\infty} dx \exp[K({\tilde \rho}) - K(x)] \left[ \frac{1}{x^2} + \frac{{\tilde \Delta}(x)}{x} \right] \ , \label{xy2} \\
&&a_1+a_2+a_3+a_4 = 2i {\tilde E}^2 {\tilde \rho} + \frac{i}{\tilde \rho} \ , \label{xy3} \\
&&a_1+a_2-a_3-a_4 = -i \int_{\tilde \rho}^{\infty} dx \exp[2K({\tilde \rho}) - 2K(x)] \left[4{\tilde E} + \frac{2\mu}{x^2} \right] - 2\int_0^{\tilde \rho} dx \exp[2K({\tilde \rho}) - 2K(x)] {\tilde E}{\tilde \Delta}(x) \ . \label{xy4}        
\end{eqnarray}
The first term in Eq. (\ref{xy3}) and the last term in Eq. (\ref{xy4}) are of third order in the small parameters ${\tilde \Delta({\tilde \rho})}$, ${\tilde E}$ and ${\tilde \rho}$ and can be neglected. The remaining two integrals can be simplified by integrating by parts
\begin{eqnarray}
&&\int_{\tilde \rho}^{\infty} dx \exp[2K({\tilde \rho}) - 2K(x)] \left[{\tilde E} + \frac{\mu}{2 x^2} \right] = - {\tilde E}{\tilde \rho} + \frac{\mu}{2 {\tilde \rho}} + \int_{\tilde \rho}^{\infty} dx \exp[2K({\tilde \rho}) - 2K(x)]\Bigl[2 {\tilde \Delta}(x) {\tilde E} x - \frac{\mu {\tilde \Delta}(x)}{x}\Bigr] \ , \nonumber \\ 
&&\int_{\tilde \rho}^{\infty} dx \exp[K({\tilde \rho}) - K(x)] \Bigl[ \frac{1}{x^2} + \frac{{\tilde \Delta}(x)}{x} \Bigr] = \frac{1}{2 {\tilde \rho}} \ . \label{byparts}
\end{eqnarray}
It is straightforward to solve the above equations for the $a_j$:
\begin{eqnarray}
&&a_1 = i\Bigl({\tilde E}{\tilde \rho} - \frac{2\mu-1}{2 {\tilde \rho}}\Bigr) - i\int_{\tilde \rho}^{\infty} dx \exp[2K({\tilde \rho}) - 2K(x)] \ {\tilde \Delta}(x) \left[2 {\tilde E} x - \frac{\mu}{x} \right] \ , \label{ax1} \\
&&a_2 = i {\tilde E}{\tilde \rho} - i\int_{\tilde \rho}^{\infty} dx \exp[2K({\tilde \rho}) - 2K(x)] \ {\tilde \Delta}(x) \left[2 {\tilde E} x - \frac{\mu}{x} \right] \ , \label{ax2} \\
&&a_3 = -i\Bigl({\tilde E}{\tilde \rho} - \frac{2\mu+1}{2 {\tilde \rho}}\Bigr) + i\int_{\tilde \rho}^{\infty} dx \exp[2K({\tilde \rho}) - 2K(x)] \ {\tilde \Delta}(x) \left[2 {\tilde E} x - \frac{\mu}{x} \right] \ , \label{ax3} \\
&&a_4 = -i {\tilde E}{\tilde \rho} + i\int_{\tilde \rho}^{\infty} dx \exp[2K({\tilde \rho}) - 2K(x)] \ {\tilde \Delta}(x) \left[2 {\tilde E} x - \frac{\mu}{x} \right] \ . \label{ax4} 
\end{eqnarray}
\end{widetext}
The common integral term in Eqs. (\ref{ax1})-(\ref{ax4}) is zero and defines the bound state energy.

\subsection{Matching of wave functions}

The solutions for ${\tilde \rho} > {\tilde \rho}_c$ are now given by $f^{\mu}_j({\tilde \rho}) = B_j [H^{(1)}_{\mu}({\tilde \rho}) g_j({\tilde \rho}) + c.c.]$, where the coefficients $B_j$ are defined at the beginning of subsection IV.D. They have to be matched at ${\tilde \rho}_c$ to $f^{\mu}_j({\tilde \rho})$ given by Eq. (\ref{Jnu}) for ${\tilde \rho} < {\tilde \rho}_c$ with the prefactors defined in subsection IV.C using a similar procedure to that of Sects. II and III and Ref. \onlinecite{CaroliDGM}. The functions $g_j$ were calculated consistently to first order of perturbation and can be written as an exponential, i.e. $g_j({\tilde \rho}) \propto \exp[-K({\tilde \rho}) + a_j({\tilde \rho})]$, which remains correct to first order. To match the wave functions we employ the asymptotic expansions for the Bessel and the Hankel functions given in Eqs. (\ref{Bessel1}) and (\ref{Hankel}).

As shown in sections II and III there are three factors in $f_j$ depending on ${\tilde \rho}$: (i) The $1/\sqrt{\tilde \rho}_c$-dependence in the Bessel and Hankel functions, (ii) the phase factors of the form $\exp[i(1 \pm {\tilde E}){\tilde \rho}_c]$, and (iii) the factors $\exp\{i[(\mu\pm1)^2-1/4]/(2{\tilde \rho}_c)\}$. We explicitly work out the matching for the function $f_1^{\mu}$; the other three functions follow similarly. For ${\tilde \rho} < {\tilde \rho}_c$ we have 
\begin{widetext}
\begin{equation}
f_1^{\mu}({\tilde \rho}_c) = A_1 J_{\mu - 1}(q_p \rho) = A_1 \sqrt{\frac{2}{\pi (1+{\tilde E}) {\tilde \rho}_c}} \cos\left[(1+{\tilde E}) {\tilde \rho}_c - \frac{\pi (\mu -1)}{2} - \frac{\pi}{4} + \frac{(\mu-1)^2-{\textstyle \frac{1}{4}}}{2(1+{\tilde E}){\tilde \rho}_c}\right] \ , \label{small1}
\end{equation}
while for ${\tilde \rho} > {\tilde \rho}_c$ we obtained
\begin{equation}
f_1^{\mu}({\tilde \rho}_c) = [H^{(1)}_{\mu}({\tilde \rho}_c) g_1({\tilde \rho}_c) + c.c.] =  \frac{1}{2}\sqrt{\frac{2}{\pi{\tilde \rho}_c}} \left\{\exp\left[i\left(\gamma + {\tilde E}{\tilde \rho}_c - \frac{(2\mu -1)}{2{\tilde \rho}_c} +{\tilde \rho}_c - \frac{\pi\mu}{2}- \frac{\pi}{4} + \frac{\mu^2-{\textstyle \frac{1}{4}}}{2{\tilde \rho}_c}\right)\right] + c.c. \right\} e^{-K({\tilde \rho}_c)} \ . \label{large1}
\end{equation}
\end{widetext}
In Eq. (\ref{large1}) the phase $\gamma$ arises from $C$, the next two terms in the exponential are due to $a_1$ and the remainder is consequence of the Hankel function.

From comparing these two expressions it follows that $A_1=\sqrt{1+{\tilde E}}$ and $\gamma = \pi/2$. As in Ref. [\onlinecite{CaroliDGM}] and Sections II and III we neglect ${\tilde E}$ in the denominator of the last term in Eq. (\ref{small1}), because ${\tilde E} \ll 1$, and the factor $e^{-K({\tilde \rho}_c)}$ in Eq. (\ref{large1}). The two expressions are then equivalent and the matching is satisfied for a large interval of ${\tilde \rho}_c$.

The above hinges on the vanishing of the common integral term in Eqs. (\ref{ax1})-(\ref{ax4}), which determines the energy of the bound state. At this point we can take ${\tilde \rho}_c \to 0$ in the lower integration limit and integrate the first term by parts:
\begin{equation}
E_{\mu} = \mu \int_0^{\infty} dx e^{-2K(x)} \frac{\Delta(x)}{x} \Big/ \int_0^{\infty} dx  e^{-2K(x)} \ . \label{bound1}
\end{equation}
Since the main contribution to the integrals is for $\rho \ll \xi$, where $\Delta(x)$ is linear in $x$, we arrive at $E_{\mu} \approx \mu \Delta'/k_F$, where $\Delta' =d\Delta/d\rho \approx k_F\Delta_{\infty}/\xi \approx k_F\Delta_{\infty}^2/E_F$ and hence
\begin{equation}
E_{\mu} \approx  \mu \frac{\Delta_{\infty}^2}{E_F}  \ . \label{bound1}
\end{equation}

Approximate expressions for the amplitudes are given by 
\begin{eqnarray}
&&f_1^{\mu}({\bf r}) = J_{\mu - 1}({\tilde \rho}) e^{-K({\tilde \rho})} e^{-i\theta} \ , \nonumber\\
&&f_2^{\mu}({\bf r}) = J_{\mu}({\tilde \rho}) e^{-K({\tilde \rho})}  \ , \nonumber \\
&&f_3^{\mu}({\bf r}) = - J_{\mu + 1}({\tilde \rho}) e^{-K({\tilde \rho})} e^{i\theta} \ , \nonumber \\
&&f_4^{\mu}({\bf r}) = J_{\mu}({\tilde \rho}) e^{-K({\tilde \rho})}  \ , \label{amplitudes1}
\end{eqnarray}
and the energy wave function with energy $E=E_{\mu}$ is then 
\begin{widetext}
\begin{equation}
{\hat \psi}_E = \int d^2r e^{-K({\tilde \rho})} \Bigl[J_{\mu - 1}({\tilde \rho}) e^{-i\theta} \psi_{\uparrow}({\bf r}) + J_{\mu}({\tilde \rho}) \psi{\downarrow}({\bf r}) - J_{\mu + 1}({\tilde \rho}) e^{i\theta} \psi^{\dagger}_{\uparrow}({\bf r}) + J_{\mu}({\tilde \rho}) \psi^{\dagger}_{\downarrow}({\bf r}) \Bigr] \ \ . \label{wfunction1}
\end{equation}
\end{widetext}
For $\mu \ne 0$ the wave function corresponds to a fermion operator with $E_{\mu} \neq 0$, while for $\mu = 0$ we have the Majorana state wave function consistent with Eq. (\ref{Mwfunction1}), up to an overall minus sign which can be absorbed into the normalization constant. 

\section{Conclusions}

We studied the bound states in the core of a vortex of a two-dimensional superconductor by solving the Bogoliubov-de Gennes equations following the procedure outlined by Caroli, de Gennes and Matricon.\cite{CaroliDGM}  For the ordinary $s$-wave superconductor we arrive at a similar result as CdeGM obtained for the 3D superconductor. The bound states are fermionic and gapped from the ground state by an energy scale of about $\Delta_{\infty}^2/2E_F$.

In Sects. III and IV the electron gas corresponds to the surface states of a topological insulator. Consequently the momentum and the spin are locked due to a strong spin-orbit interaction. Two cases have been considered, namely, a locking of the spin parallel and perpendicular to the momentum. The superconductivity is induced into the 2D Dirac sea via proximity of an $s$-wave superconductor.\cite{DasSarma1,DasSarma2} The results for the bound states in the core of the vortex are independent of the kind of spin-orbit coupling (as long as it is strong). In the Appendix we present the unitary transformation connecting the Hamiltonians for parallel and perpendicular spin and momentum locking (spin-rotation). The characteristic energy scale for the spacing of the energy levels is $\Delta_{\infty}^2/E_F$. 

The calculation yields a string of fermion bound states with energy $E_{\mu}$, $\mu \neq 0$ and a bound state with Majorana statistics for $\mu=0$ and $E=0$. We obtained analytical expressions for the energy spectrum and the wave functions. The main difference between the ordinary superconductor and the topological superconducting gas is the spin-locking. In the latter in a closed path the spin is forced to follow the momentum giving rise to a non-trivial Berry phase of 1/2. This converts the half-integer quantum numbers into integer ones and opens the possibility to the existence of a Majorana fermion. 

Within the range of validity of the present calculation ($|E| \ll \Delta_{\infty} \ll E_F$), the gap between the Majorana state and the first excited fermion state is very small. Hence, extremely low temperatures are required, unless $E_F$ is reduced to close to the vertex of the Dirac Hamiltonian.  Although this is beyond the validity of our results, we do not expect qualitative changes in the results. Indeed it has been numerically shown in Ref. [\onlinecite{DasSarma1}] that the first excited state above the Majorana bound state can have an excitation energy of $\Delta_{\infty}$. This would be a necessary condition for the use of this Majorana state in quantum computing.

\vskip 0.2in
\acknowledgements
The authors thank Prof. P. Xiong for helpful comments. The support by the U.S. Defense Advanced Research Projects Agency (DARPA) under agreement number D18AC00010 is acknowledged. The National High Magnetic Field Laboratory is supported by National Science Foundation Cooperative Agreement No. DMR1644779 and the State of Florida.
\vskip 0.2in
{\bf Appendix: Spin rotation transformation}
\vskip 0.2in
In this Appendix we present the unitary transformation mapping the Hamiltonian with parallel spin-momentum locking (section III), ${\check {\cal H}_B^{\parallel}}$ (Eq. (\ref{h})), onto the model with Rashba interaction (perpendicular coupling, section IV), ${\check {\cal H}_B^{\perp}}$ (Eq. (\ref{h1})). Consider the four-dimensional unitary matrix
\begin{eqnarray}
U &=& \left[ \begin{array}{cc} e^{i(\pi/4)\sigma_z} & 0 \\ 0 & e^{-i(\pi/4)\sigma_z} \end{array} \right] \nonumber \\ 
&=& \ \left[ \begin{array}{cc} (1 + i \sigma_z)/\sqrt{2} & 0 \\ 0 & (1 - i \sigma_z)/\sqrt{2} \end{array} \right] \ . \label{transf} 
\end{eqnarray}
It is now straightforward to show that ${\check {\cal H}}_B^{\perp} = U {\check {\cal H}}_B^{\parallel} U^{\dagger}$, 
\begin{widetext}
\begin{eqnarray}
U{\check {\cal H}}_B^{\parallel} U^{\dagger} &=& \left[ \begin{array}{cc} e^{i(\pi/4)\sigma_z} & 0 \\ 0 & e^{-i(\pi/4)\sigma_z} \end{array} \right] \left[ \begin{array}{cc} \sigma_x p_x +\sigma_y p_y - E_F & i \sigma_y \Delta \\ - i \sigma_y \Delta^* & -(\sigma_x p_x +\sigma_y p_y)^* + E_F \end{array} \right] \left[ \begin{array}{cc} e^{-i(\pi/4)\sigma_z} & 0 \\ 0 & e^{i(\pi/4)\sigma_z} \end{array} \right] \nonumber \\
&=& \left[ \begin{array}{cc} (-\sigma_y p_x +\sigma_x p_y) - E_F & i \sigma_y \Delta \\ -i \sigma_y \Delta^* & -(-\sigma_y p_x +\sigma_x p_y)^* + E_F \end{array} \right] = {\check {\cal H}}_B^{\perp} \ . \label{transf1}
\end{eqnarray}
The wave functions, including the phase factors $B_j$, for the two types of spin-momentum lockings then also transform accordingly.

\end{widetext}

\end{document}